\documentclass[aps,12pt,preprint,groupedaddress,showpacs,reprintnumbers,amsmath,amssymb,floatfix]{revtex4}
\usepackage{graphicx}
\usepackage{epsfig}
\usepackage{graphics}
\usepackage{latexsym}
\usepackage{rotating}
\usepackage{bm}
\usepackage{color}
\usepackage[T1]{fontenc}
\usepackage[latin9]{inputenc} 
\usepackage{color}
\usepackage{esint}
\usepackage{url}
\global\long\def\k{\mathbf{k}}

\global\long\def\kp{\mathbf{k'}}

\global\long\def\q{\mathbf{q}}

\global\long\def\qp{\mathbf{q'}}

\global\long\def\r{\mathbf{r}}

\begin{document}

%\preprint{ADP-THY-21/xxxx}

\title{Hartree-Fock Formulation of the QMC Model at Finite Temperature}

\author{P. A. M. Guichon}
\affiliation{DPhN, IRFU-CEA, Universit\'e Paris-Saclay, F-91191 Gif sur Yvette, France}
\author{J.~R.~Stone}
\affiliation {Department of Physics and Astronomy, University of Tennessee, Knoxville, TN 37996, USA\\
Department of Physics (Astrophysics), University of Oxford, Oxford OX1 3RH, United Kingdom}
\author{A. W. Thomas}
\affiliation{CSSM and ARC Centre of Excellence for Dark Matter Particle Physics, Department of Physics, School of Physical Sciences, University of Adelaide, Adelaide SA 5005, Australia}

\date{\today}

\begin{abstract}
We present, for the first time, a detailed theory of high density matter including the entire baryon octet at finite temperature, based on a fully relativistic mean field model with a \emph{consistent treatment of exchange (Fock) terms}, using the quark-meson-coupling model (QMC). It has been already demonstrated that the QMC equation of state is applicable in thermodynamic scenarios in stationary and rotating isentropic proto-neutron stars, producing results in agreement with recent observation. It is also suitable for the simulation of the behaviour following a binary neutron star merger~\cite{Stone2021}; \url{https://compose.obspm.fr/eos/205}. 

We develop a comprehensive demonstration of the impact of the Fock
terms in the QMC energy density functional on properties of neutrinoless proto-neutron stars with cores containing the full hyperon octet with constant entropy, S/A=2k$_B$. 
Given the interest in the properties of the proto-neutron star remaining after either a supernova explosion or the merger of two neutron stars, it is vital to develop modern equations of state at finite temperature. While much attention has been paid to relativistic mean-field calculations at finite temperature, it is crucial to explore the consequences of a consistent treatment of the Fock terms. 
\end{abstract}

\pacs{}

\maketitle

\section{Introduction}

The observation of gravitational waves from binary neutron star (BNS) mergers by the LIGO and Virgo collaborations~\cite{LIGOScientific:2017vwq} has generated considerable scientific interest. Neutron stars contain the most dense matter in the Universe and new information of this kind~\cite{LIGOScientific:2018cki} is potentially extremely valuable in the quest to understand the equation of state (EoS) and the very nature of such dense matter~\cite{Bauswein:2018bma,Blacker:2020nlq,Weih:2019xvw,Liebling:2020dhf,Constantinou:2021hba}. As just one example, within the community working on hadronic EoS there is tremendous interest in the role of hyperons~\cite{Sekiguchi:2011mc,Radice:2016rys,Blacker:2023opp}. Studies of the dynamics of BNS mergers within General Relativity suggest~\cite{Endrizzi:2019trv,Kashyap:2021wzs,Issifu:2023qyi} that the matter existing in the first 20 msec after the merger will experience temperatures of order 10-20 MeV, and possibly considerably higher. This is a period when observed gravitational wave signals are generated and so it is vital that one can make available EoS at these temperatures with hyperons included. 

In a previous paper~\cite{Stone2021} we examined high-density matter in cores of cold neutron stars (NS) and hot isentropic proto-neutron stars (PNS) using the quark-meson-coupling model (QMC-A)~\cite{Guichon2018} extended to finite temperatures. The temperature effects were demonstrated in two scenarios, (i) lepton rich matter with trapped neutrinos and lepton fraction Y$_L$=0.4 and entropy S/A=1k$_B$  and (ii) deleptonized,  chemically equilibrated matter with 
S/A=2k$_B$, both containing either only nucleons or the full baryon octet. The EoS, gravitational mass, radius, baryon composition, moments of inertia and Kepler frequency for slow and fast rigidly rotating stars were explored over a wide range of temperatures and baryon number densities. The nucleon-hyperon phase transition was studied through the adiabatic index and the speed of sound. The results were compared with two relativistic mean field (RMF) models, the chiral mean field model (CMF) \cite{Papazoglou:1998vr,Dexheimer2008,Roark2019,Dexheimer:2018dhb} and the generalized relativistic density functional (GRDF) with DD2 (nucleon-only) and DD2Y-T (full baryon octet) 
interactions~\cite{Typel2010,Pais2017,Marques2017}. Full EoS tables, covering the range of temperatures from T=0 to 100 MeV, entropy per particle (S/A) between 0 and 6, lepton fraction from Y$_L$=0.0 to 0.6, and baryon number density range n$_B$=0.05-1.2 fm$^{-3}$, suitable either for simulations of core-collapse supernova (CCSN) or  NS merger modelling, have been posted at \url{https://compose.obspm.fr/eos/205,  https://compose.obspm.fr/eos/206)}.

However, the full details of the general QMC derivation at finite temperature have not been published yet. Here we present, for the first time, a fully consistent, relativistic Hartree-Fock formulation of the EoS of hyperonic matter at high density and temperature. There have been a number of mean-field simulations of PNS properties in the literature, for example~\cite{Pons1999,Dexheimer2008,Sumiyoshi2009,Panda2010,Ishizuka2008,Oertel2012,Marques2017}, but almost all have used mean-field approximation, omitting the exchange, Fock terms. The single exception of which we are aware that included the Fock terms was applied to matter containing only nucleons~\cite{Zhang2016}.

The main objectives of this work are to report the theoretical development of the QMC model for hot proto-neutron stars,  including the effect of the Fock terms, and to
present a comprehensive demonstration of the impact of these Fock terms on the properties of neutrinoless PNS, with
cores containing the full hyperon octet. In order to explore these effects we have chosen to work at constant entropy, S/A=
2k$_B$. While this system has been chosen as an example, consistent with our previous investigation~\cite{Stone2021,Stone2022c}, it has a more general significance representing the state of hyperonic matter just after deleptonization. 

In Sec.~\ref{sec:qmc} we present the Hamiltonian  of the quark-meson coupling model and its thermodynamics. We explain how we compute the partition function in   Sec.~\ref{sec:Finite-temp-HF}. In
Sec.~\ref{sec:thermo} we give explicit expressions for relevant
thermodynamic quantities, as well as the conditions for chemical
equilibrium. The computational details are presented in Sec.~\ref{sec:com}, while the results and discussion form the content of
Sec.~\ref{sec:resdis}. Concluding remarks are presented in Sec.~\ref{sec:concl}.

%%%%%%%%%%%%%%%%%%%%%%%%%%%%%%%%%%%%%%%%%%%%%%%%%%%%%%%%%%%%%%%%%
\section{The quark-meson coupling model}
\label{sec:qmc}

\subsection{Hamiltonian}

In order to ensure that the presentation is as accessible as possible we begin 
with the simplest version of the quark-meson coupling (QMC) model. This model takes into account the effect of the exceptionally strong relativistic mean scalar fields in dense nuclear matter (see for example Ref.~\cite{Brockmann1990}) upon the internal structure of the bound hadrons~\cite{Guichon1988,Guichon1996,Saito2007,Guichon2018}. This leads naturally to the introduction of the scalar polarizability, which describes the fact that the internal valence quark wave functions adjust self-consistently to oppose the applied scalar field, just as, for example, the electric polarizability opposes an applied electric field. The density dependence introduced in this way is equivalent to introducing repulsive three-body forces between the hadrons in the medium~\cite{Guichon2004,Guichon2006} with no additional parameters. Indeed, as the Lorentz scalar and vector interactions between the hadrons are generated by the exchange of mesons between the confined quarks in different hadrons (e.g., the $\sigma$ meson for the scalar-isoscalar force and the $\omega$ for the vector-isoscalar force), these many-body forces are entirely determined by the particular confining quark model under consideration~\cite{Bentz:2001vc,Whittenbury:2015ziz}. 

To introduce the finite temperature formalism, we first consider just the $\sigma$ and $\omega$ mesons interacting with a single flavor of fermion. 
In this case the Hamiltonian takes the form~\cite{Guichon1988}
\[
H=H^{B}+H^{M} \, ,
\]
where the meson part, which we assume static, is:
\[
H^{M}=\int d\vec{r}\left[\frac{1}{2}\left(\vec{\nabla}\sigma\right)^{2}+V(\sigma)\right]-\frac{1}{2}\int d\vec{r}\left[\left(\vec{\nabla}\omega\right)^{2}+m_{\omega}^{2}\omega^{2}\right]
\, . \] 
The $\sigma$ potential is taken to be
\begin{equation}
V(\sigma)=\frac{1}{2}m_{\sigma}^{2}\sigma^{2}+\frac{\lambda_{3}}{3!}\left(g_{\sigma}\sigma\right)^{3}+\frac{\lambda_{4}}{4!}\left(g_{\sigma}\sigma\right)^{4} \, ,
\label{eq:Vsigma}
\end{equation}
and in practice we set $\lambda_{4}=0$.
Note that we neglect the  spatial components of the meson  fields because their expectation value vanishes in uniform matter.

The baryon component of the Hamiltonian in the finite volume ${\cal V}$ is 
\[
H^{B}=\frac{1}{{\cal V}}\int d^{3}\mathbf{r}\sum_{\mathbf{kk'}}e^{i(\mathbf{k-k'}).\mathbf{r}}a_{\mathbf{k}}^{\dagger}a_{\mathbf{k'}}K(\k,\kp,\sigma,\omega)
\, , \] 
where the sum includes the sum over spin. The infinite volume limit amounts to the replacement 

\[
\frac{1}{{\cal V}}\sum_{\mathbf{k}}\to\frac{2}{(2\pi)^{3}}\int d\mathbf{k}
\]
The kinetic  term is defined as
\[
K(\k,\kp,\sigma,\omega)=\frac{1}{2}\left(\sqrt{\mathbf{k}^{2}+M(\sigma)^{2}}+\sqrt{\mathbf{k'}^{2}+M(\sigma)^{2}}\right)+g_{\omega}\omega
\]
with the effective mass
\begin{equation}
M(\sigma)=M-g_{\sigma}\sigma+\frac{d}{2}\left(g_{\sigma}\sigma\right)^{2} 
\label{eq:Msigma} \, .
\end{equation}
and we define $K(\k,\sigma,\omega) \equiv K(\k,\k,\sigma,\omega)$. 

The coupling constants $g_{\sigma}$ and $g_{\omega}$ are, respectively, the couplings of the $\sigma$ and $\omega$ mesons to the nucleon in free space, which in turn are calculated in terms of the more fundamental couplings to the $u$ and $d$ quarks confined in the MIT bag~\cite{DeGrand1975}.  The effect of the self-consistent solution of the coupling of the scalar meson to the confined quarks is reflected in the scalar polarizability, $d$, appearing in Eq.~(\ref{eq:Msigma}). This is not a free parameter but must be calculated within the particular confining quark model under consideration. The generalisation to include the effect of  more than one flavor of baryon flavor is given in Refs.~\cite{Guichon:2008zz,Guichon2018}.

%%%%%%%%%%%%%%%%%%%%%%%%%%%%%%%%%%%%%%%%%%%%%%%%%%%%%%

\subsection{Thermodynamics of the model}

To determine the thermodynamic properties of nuclear matter at temperature
$T$ we use the grand cannonical ensemble. We compute the partition
function 
\begin{equation}
Z({\cal V},\beta,\mu)=Tr\,e^{-\beta(H-\mu N)}=e^{-\beta\Phi}\label{eq:Z-0}
\end{equation}
where ${\cal V}$ is the volume, $\beta=1/kT$ and $\mu=\{\mu(p),\mu(n),\ldots\}$
stands collectively for the chemical potentials of the baryons. Here
we assume that only members of the baryon octet,
$\{p,n,\Lambda,\cdots\}$, 
are present in the system. In Eq.(\ref{eq:Z-0}) $\Phi$ is the
grand potential, $H$ the total Hamiltonian and $N$ stands collectively
for the particle numbers of the various baryon flavors. In the QMC model~\cite{Guichon2018}
$H$ depends on the second quantized operators $a_{k,f}$ of the baryons
and on the fields $(\sigma,\omega)$, which
describe the $\sigma$ and $\omega$ mesons. The trace in Eq.~(\ref{eq:Z-0})
involves both a sum over baryon states and a functional integration
over the meson fields, which are time independent in the model. 

We assume that the meson fields can each be written as a C-number, 
$\bar{\sigma},\bar{\omega}$, plus small fluctuations, so that:
\begin{eqnarray*}
\sigma & = & \bar{\sigma}+\sum_{\q\neq0}e^{i\q.\r}\delta\sigma_{\q}\\
\omega & = & \bar{\omega}+\sum_{\q\neq0}e^{i\q.\r}\delta\omega_{\q} \, , 
\end{eqnarray*}
where the zero mode is excluded from the sum and we impose $\delta\sigma_{\q}=\delta\sigma_{-\q},\,\delta\omega_{\q}=\delta\omega_{-\q}$
to ensure that the fields are Hermitian. The integration over $\sigma,\omega$ is then (up to an irrelevant multiplicative factor) 
\[
\int D\sigma D\omega=\int d\bar{\sigma}d\bar{\omega}\int\prod_{\q}\delta\sigma_{\q}\prod_{\qp}\delta\omega_{\qp} \, .
\]
Since the thermodynamic functions involve only the logarithmic derivatives 
with respect to $\beta,\mu,{\cal V}$ , multiplicative factors which are independent
of these variables can be ignored.

We expand the Hamiltonian up to terms quadratic in the fluctuations. For
the meson part we find:
\begin{eqnarray*}
H^{M} & = & {\cal V}\left[V(\bar{\sigma})+\frac{1}{2}\sum_{\q}\delta\sigma_{\q}^{2}\left(\q^{2}+\frac{d^{2}V}{d\sigma^{2}}\right)\right]\\
 & - & {\cal V}\left[\frac{m_{\omega}^{2}}{2}\bar{\omega}^{2}+\frac{1}{2}\sum_{\q}\delta\omega_{\q}^{2}\left(\q^{2}+m_{\omega}^{2}\right)\right]
\, ,
\end{eqnarray*}
while the baryon part becomes:
\[
H^{B}=H_{0}+\sum_{\q\neq0} \left( H_{\q}^{\sigma}\delta\sigma_{\q}+H_{\q}^{\omega}\delta\omega_{\q} \right) +\sum_{\q\qp\neq0}H_{\q-\qp}^{m}\delta\sigma_{\q}\delta\sigma_{\qp} 
 \, , \]
with
\begin{eqnarray*}
H_{0} & = & \sum_{\k}K^{\k\k}(\bar{\sigma},\bar{\omega})a_{\k}^{\dagger}a_{\k}\\
H_{\q}^{\sigma} & = & \frac{1}{2}\sum_{\k}\left(a_{\k}^{\dagger}a_{\k+\q}\frac{\partial K^{\k\k+\q}}{\partial\sigma}+\q\to-\q\right),\\
H_{\q\qp}^{m} & = & \frac{1}{2}\sum_{\k}a_{\k}^{\dagger}a_{\k+\q+\qp}\frac{\partial^{2}K^{\k\k+\q+\qp}}{\partial\sigma^{2}},\\
H_{\q}^{\omega} & = & \frac{1}{2}\sum_{\k}\left(a_{\k}^{\dagger}a_{\k+\q}\frac{\partial K^{\k\k+\q}}{\partial\omega}+\q\to-\q\right).
\end{eqnarray*}
Here we have used the symmetry $\delta\sigma_{-q}=\delta\sigma_{q}$, so
that $H_{\q}^{\sigma}=H_{-\q}^{\sigma}$ and 
$H_{\q}^{\sigma\dagger}=H_{\q}^{\sigma}$ and similarly for the component of the Hamiltonian involving the $\omega$ field.

The integration over $\delta\sigma_{\q},\delta\omega_{\q}$ can be
carried out explicitly, since the dependence on these fluctuations is quadratic  
\begin{eqnarray*}
H-\mu N & = & {\cal V}\left(V(\bar{\sigma})-\frac{m_{\omega}^{2}\bar{\omega}}{2}\right)+\sum_{\k}K^{\k\k}(\bar{\sigma},\bar{\omega})a_{\k}^{\dagger}a_{\k}-\mu N\\
 & + & \frac{{\cal V}}{2}\left(\sum_{\q\ne0}\delta\sigma_{\q}^{2}\left(\q^{2}+\frac{d^{2}V}{d\sigma^{2}}\right)-\sum_{\q\ne0}\delta\omega_{\q}^{2}\left(\q^{2}+m_{\omega}^{2}\right)\right)\\
 & + & \sum_{\q\ne0}\left(\tilde{H}_{\q}^{\sigma}\delta\sigma_{\q}+\tilde{H}_{\q}^{\omega}\delta\omega_{\q}\right)+\sum_{\q\qp\ne0}H_{\q-\qp}^{m}\delta\sigma_{\q}\delta\sigma_{\qp} \, . 
\end{eqnarray*}
The contribution arising from the $\delta\sigma$ integration can be written
as
\[
Z_{\delta\sigma}=\int\prod_{\q}\delta\sigma_{\q}e^{-\beta S}
\, , \]
with
\[
S=\frac{{\cal V}}{2}\sum_{\q\ne0}\delta\sigma_{\q}^{2}\left(\q^{2}+\frac{d^{2}V}{d\sigma^{2}}\right)+\sum_{\q\qp\ne0}H_{\q-\qp}^{m}\delta\sigma_{\q}\delta\sigma_{\qp}+\sum_{\q\ne0}H_{\q}^{\sigma}\delta\sigma_{\q} \, .
\, . \]
By a change of integration variables (which induces only an irrelevant multiplicative Jacobian) we can choose $\delta\sigma_{\q}$ so as to diagonalize the quadratic part of $S$, that is
\begin{equation}
\sum_{\qp}\left[\frac{{\cal V}}{2}\left(\q^{2}+\frac{d^{2}V}{d\sigma^{2}}\right)\delta(\q,\qp)+H_{\q-\qp}^{m}\right]\delta\sigma_{\qp}=\alpha(\q)\delta\sigma_{\q} \, .
\label{eq:diagonalize}
\end{equation}
So we can write
\[
S=\sum_{\q\ne0}\alpha(\q)\delta\sigma_{\q}^{2}+H_{\q}^{\sigma}\delta\sigma_{\q}=\sum_{\q\ne0}\left[\sqrt{\alpha(\q)}\delta\sigma_{\q}+\frac{H_{\q}^{\sigma}}{2\sqrt{(\alpha(\q)}}\right]^{2}-\sum_{\q\ne0}\left(\frac{H_{\q}^{\sigma}}{2\sqrt{(\alpha(\q)}}\right)^{2} \, .
\]
The first term in $S$ contributes an irrelevant factor to $Z$, so the
contribution of the $\delta\sigma$ integration is simply
\[
Z_{\delta\sigma}=e^{-\beta H_{\delta\sigma}} \, ,
\]  
 with 
\[
H_{\delta\sigma}=-\sum_{\q\ne0}\left(\frac{H_{\q}^{\sigma}}{2\sqrt{(\alpha(\q)}}\right)^{2} \, .
\]

We estimate the effect of $H^{m}$ in Eq.~(\ref{eq:diagonalize}) assuming
it is a perturbation. At leading order one finds
\begin{equation}
\alpha(q)\sim\frac{{\cal V}}{2}\left(q^{2}+\frac{d^{2}V}{d\sigma^{2}}\right)+H_{q-q}=\frac{{\cal V}}{2}\left(q^{2}+\frac{d^{2}V}{d\sigma^{2}}+\frac{1}{{\cal V}}\sum a_{k}^{\dagger}a_{k}\frac{\partial^{2}K^{kk}}{\partial\sigma^{2}}\right) \, .
\label{eq:alpha}
\end{equation}
Obviously this induces an effective $\sigma$ mass:
\begin{eqnarray*}
\tilde{m}_{\sigma}^{2} & = & m_{\sigma}^{2}+\lambda_{3}\bar{\sigma}+\frac{1}{{\cal V}}\sum a_{k}^{\dagger}a_{k}
\frac{\partial^{2}K^{kk}}{\partial\sigma^{2}} \, .
\end{eqnarray*}
In the following we neglect this effect because
the bare $\sigma$ mass is not so well known. So, keeping the leading
term in Eq.~\ref{eq:alpha}, we obtain
\[
H_{\delta\sigma}=-\frac{1}{2{\cal V}}\sum_{\q\ne0}\frac{\left(H_{\q}^{\sigma}\right)^{2}}{\q^{2}+m_{\sigma}^{2}} \, . 
\]  

In summary, we have: 
\begin{equation}
Z=\int d\bar{\sigma}d\bar{\omega}\sum_{n}\langle n|e^{-\beta(H_{mean}+H_{fluc}-\mu N)}|n\rangle \, ,
\label{eq:Z_fluct}
\end{equation}
so that, including the $\omega$ by analogy: 
\begin{eqnarray}
H_{mean} & = & E_{meson}+H_{0}\nonumber \\
H_{0} & = & \sum_{\k}K^{\k\k}(\bar{\sigma},\bar{\omega})a_{\k}^{\dagger}a_{\k}\nonumber \\
E_{meson} & = & {\cal V}\left(V(\bar{\sigma})-\frac{m_{\omega}^{2}\bar{\omega}^{2}}{2}\right)\nonumber \\
H_{fluc} & = & -\frac{1}{2{\cal V}}\sum_{\q\neq0}\frac{\left(H_{\q}^{\sigma}\right)^{2}}{\q^{2}+m_{\sigma}^{2}}+\frac{1}{2{\cal V}}\sum_{\q}\frac{\left(H_{\q}^{\omega}\right)^{2}}{q^{2}+m_{\omega}^{2}} \, .
\label{eq:Hfluctu}
\end{eqnarray}
The generalization to include flavor and isovector exchange is given below.

%%%%%%%%%%%%%%%%%%%%%%%%%%%%%%%%%%%%%%%%%%%%%%%%%%%%%%%%%%%
\section{Finite temperature Hartree Fock method with hyperons}
\label{sec:Finite-temp-HF} 

\subsection{Perturbative effect of $H_{fluc}$}
The next step is to compute the sum over hadronic states in 
Eq.~(\ref{eq:Z_fluct}).
We write 
\begin{eqnarray*}
H_{0}+H_{fluc} & = & \sum_{\k}e(k)a_{\k}^{\dagger}a_{\k}+\delta H\\
\delta H & = & \sum_{\k}\left[K(\k)-e(\k)\right]a_{\k}^{\dagger}a_{\k}+H_{fluc} \, ,
\end{eqnarray*}
and assume that one can choose $e(k)$ such that $\delta H$ can
be considered as a perturbation. Using thermal perturbation theory~\cite{Negele2018} at leading order in $\delta H$ we get, 
with 
\[Z=\exp-\beta\left(\Phi_{B}+E_{meson}\right) \, , \]
where
\begin{eqnarray*}
\Phi_{B} & = & \Phi_{B0}+\sum_{\k}n(\k)\left[K(\k)-e(\k)\right]+\\
 &  & \frac{1}{4{\cal V}}\sum_{\k\kp}n(\k) \, n(\kp)\left[\frac{1}{\left(\k-\kp\right)^{2}+m_{\sigma}^{2}}\left(\frac{\partial K^{kk'}}{\partial\sigma}\right)^{2}-\frac{1}{\left(\k-\kp\right)^{2}+m_{\omega}^{2}} \, g_{\omega}^{2}\right]
\end{eqnarray*}
and $\Phi_{B0}$ is defined in Eq.~(\ref{eq:Phi-generic}).

We now introduce the flavor dependence, along with the isovector interaction
associated with $\rho$ and $\pi$ exchange. The mean field corresponding to the time component of the neutral $\rho$ meson is labelled $b_3$, in order to distinguish it from the density, $\rho$. In the infinite volume limit this leads to: 
\begin{eqnarray}
\Phi_{B} & = & \Phi_{B0}+\frac{2{\cal V}}{(2\pi)^{3}}\sum_{f}\int d\vec{k} \, n(k,f)\left[K(k,f)-e(k,f)\right]\nonumber \\
 & + & \frac{{\cal V}}{(2\pi)^{6}}\sum_{ff'}\int d\vec{k} \, d\vec{q} \, n(k,f) \, n(q,f') \, W^{ff'}(\vec{k},\vec{q})
\label{eq:Phi-generic}\\
\Phi_{B0} & = & -\frac{{\cal V}}{\beta}\frac{2}{(2\pi)^{3}}\sum_{f}\int d\vec{k} \, \ln\left(1+e^{-\beta\left(e(k,f)-\mu(f)\right)}\right) \, ,
\nonumber 
\end{eqnarray}
with
\[
n(k,f)=\frac{1}{1+\exp\left[\beta(e(k,f)-\mu(f)\right]} \, ,
\]
and 
\begin{eqnarray}
K(k,f) & = & \sqrt{k^{2}+M_{f}^{2}(\bar{\sigma})}+g_{\omega}^{f}\bar{\omega}+g_{\rho}m(f)\bar{b}_{3}\label{eq:K}\\
E_{meson} & = & {\cal V}\left(V(\bar{\sigma})-\frac{m_{\omega}^{2} \, \bar{\omega}^{2}}{2}-\frac{m_{\rho}^{2} \, \bar{b}_{3}^{3}}{2}
\right) \, ,
\label{eq:Emeson}
\end{eqnarray}
where $m(f)$ is the isospin projection of the flavor $f=p,n,\Lambda,...$ ($ + \frac{1}{2}, \, -\frac{1}{2}, \, 0 \, ....$).

The kernel $W^{ff'}(k,q)$ describes the Fock terms associated with the four mesons
$(\sigma,\omega,\rho,\pi)$ :
\begin{eqnarray*}
W^{ff'}(\vec{k},\vec{q}) & = & \delta(f,f')\left[\frac{1}{(\vec{k}-\vec{q})^{2}+m_{\sigma}^{2}}\left(\frac{\partial K_{f}^{kq}}{\partial\sigma}\right)^{2}-\frac{1}{\left(\vec{k}-\vec{q}\right)^{2}+m_{\omega}^{2}}\left(g_{\omega}^{f}\right)^{2}\right]\\
 &  & -G_{\rho}S(f,f')\frac{m_{\rho}^{2}}{(\vec{k}-\vec{q})^{2}+m_{\rho}^{2}}-\left(\frac{g_{A}}{2f_{\pi}}\right)^{2}\Pi(f,f')\frac{m_{\pi}^{2}}{(\vec{k}-\vec{q})^{2}+m_{\pi}^{2}} \, .
\end{eqnarray*}
Here we have defined
\begin{eqnarray}
S_{ff'} & = & \delta_{mm'}m^{2}+t(\delta_{m,m'+1}+
\delta_{m',m+1}) \, ,
\label{eq:rho_coeff}
\end{eqnarray}
where $(t,m)$ are the isospin labels corresponding to the baryon of flavor $f$ and the  matrix $\Pi(f,f')$ is:
\[
\begin{array}{ccccccccc}
 & p \, & n & \Lambda & \Sigma_{-} & \Sigma_{0} & \Sigma_{+} & \Xi_{-} & \Xi_{0}\\
p & 1 \, & 2 & 0 & 0 & 0 & 0 & 0 & 0\\
n & 2 \, & 1 & 0 & 0 & 0 & 0 & 0 & 0\\
\Lambda & 0 \, & 0 & 0 & -12/25 & -12/25 & -12/25 & 0 & 0\\
\Sigma_{-} & 0 \, & 0 & -12/25 & 16/25 & 16/25 & 0 & 0 & 0\\
\Sigma_{0} & 0 \, & 0 & -12/25 & 16/25 & 0 & 16/25 & 0 & 0\\
\Sigma_{+} & 0 \, & 0 & -12/25 & 0 & 16/25 & 16/25 & 0 & 0\\
\Xi_{-} & 0 \, & 0 & 0 & 0 & 0 & 0 & 1/25 & 2/25\\
\Xi_{0} & 0 \, & 0 & 0 & 0 & 0 & 0 & 2/25 & 1/25
\end{array}
\]
%

%%%%%%%%%%%%%%%%%%%%%%%%%%%%%%%%%%%%%%%%%%%%
\subsection{The determination of $e(k,f)$}
We determine $e(k,f)$ by applying the finite temperature Hartree-Fock variational principle 
\begin{equation}
\frac{\delta\Phi}{\delta e(p,m)}=0 \, .
\label{eq:HF_condition}
\end{equation}
This condition is applied for arbitrary values of 
$\bar{\sigma},\bar{\omega},\bar{b}_{3}$. 
Then, using Eq.~(\ref{eq:Phi-generic}), we find
\begin{eqnarray*}
\frac{\delta\Phi}{\delta e(p,m)} & = & \frac{2}{(2\pi)^{3}}\sum_{f}\int d\vec{k}\frac{\delta n(k,f)}{\delta e(p,m)}\\
 &  & \left[K(k,f)-e(k,f)+\frac{1}{(2\pi)^{3}}\sum_{f'}\int d\vec{q} \, n(q,f') \, W^{ff'}(\vec{k},\vec{q})\right] \, .
\end{eqnarray*}
Hence the Hartree-Fock (HF) equations are
\begin{eqnarray}
e(k,f) & = & K(k,f)+\frac{1}{(2\pi)^{3}}\sum_{f'}\int d\vec{q} \, n(q,f') \, W^{ff'}(k\hat{z},\vec{q}) \, ,
\label{HF-eqs}
\end{eqnarray}
where the rotational invariance of $W$ has been used to put $\vec{k}$
along an arbitrary axis. This clearly generalizes the HF equations
for the single particle energies $e(k,f)$. We note that these equations
must be solved self-consistently, because $n(k,f)$ depends on $e(k,f)$. If we substitute 
Eq.~(\ref{HF-eqs}) into the expression for the grand potential we find:
\begin{eqnarray}
\Phi_{B} & = & \Phi_{B0}-\frac{{\cal V}}{(2\pi)^{6}}\sum_{ff'}\int d\vec{k} \, d\vec{q} \, n(k,f) \, n(q,f') \,W^{ff'}(\vec{k},\vec{q})\label{Phi_B1}\\
 & = & \Phi_{B0}+\frac{{\cal V}}{(2\pi)^{3}}\sum_{f}\int d\vec{k} \, n(k,f)	 \left[K(k,f)-e(k,f)\right] \, .
\label{Phi_B2}
\end{eqnarray}
Note that the above expressions, 
Eqs.~(\ref{Phi_B1},\ref{Phi_B2}), which
give the values of $\Phi_{B}$ at the solution, are not stationary
with respect to variation of $e(k,f)$. If one needs to invoke the
stationarity one must use the full expression, 
Eq.~(\ref{eq:Phi-generic}).

%%%%%%%%%%%%%%%%%%%%%%%%%%%%%%%%%%%%%%%%%%%%%%%
\subsection{Solving the Hartree-Fock equation}
For clarity of presentation we omit the dependence on unnecessary parameters and write
the self-consistent Eqs.~(\ref{HF-eqs}) in the symbolic form
\[
e(T)=K+F\left[e(T),T\right] \, .
\]
We assume that we can solve them by iteration:
\begin{equation}
e^{(n+1)}(T)=K+F\left[e^{(n)}(T),T\right] \, ,
\label{eq:iteration}
\end{equation}
realizing that the choice of the initial step, $e^{(0)}(T)$, is critical. In the initial work in  Ref.~\cite{Stone2021}, we used the obvious choice $e^{(0)}(T)=K$, which resulted in very slow convergence as $T \to 0$, because the Fermi distribution
becomes singular in this limit. To obtain satisfactory convergence
we introduced a form factor to cut off the high momenta. In
practice we made the replacement
\[
W^{ff'}(\vec{k},\vec{q})\to W^{ff'}(\vec{k},\vec{q})\left(\frac{M_{r}^{2}}{(M_{r}^{2}+(\vec{k}-\vec{q})^{2}}\right)^{2} \, ,
\]
with $M_{r}\sim0.5\div1GeV.$ Although this can be interpreted as an effect of the hadron size, it may also be viewed as an ad hoc recipe.

Fortunately there is a much better solution. Suppose that $e(T)$ is
the exact solution at some temperature $T$. Then if we choose the
starting point 
\[
e^{(0)}(T+\Delta T)=e(T) \, ,
\]
we have a good chance that the iteration process at $T+\Delta T$
will converge rapidly, provided that $\Delta T$ is not too large.

Of course, we do not have the solution at finite $T$ but at least we can find it relatively easily at $T=0.$ Indeed, at $T=0$ 
Eqs.~(\ref{HF-eqs}) become
\begin{eqnarray}
e(k,f) & = & K(k,f)+\frac{1}{(2\pi)^{3}}\sum_{f'}\int_{0}^{k_{F}(f')}d\vec{q} \,\, W^{ff'}(k\hat{z},\vec{q}) \, ,
\label{HF-eqs-1}
\end{eqnarray}
with the Fermi momentum defined by 
\[
e(k_{F}(f),f)=\mu(f) \, .
\]
If we substitute this into Eq.~(\ref{HF-eqs-1}) we get 
\[
\mu(f)=K(k_{F}(f),f)+\frac{1}{(2\pi)^{3}}\sum_{f'}\int_{0}^{k_{F}(f')}d\vec{q} \,\,W^{ff'}(k_{F}\hat{z},\vec{q}) \, .
\]
These equations determine $k_{F}(f)$ (or equivalently the density of flavor $f$), 
when the chemical potentials $\mu(f)$ are given. This allows one to pass
from the grand canonical to the canonical ensemble. For our purpose
what matters is that the equations for $k_{F}(f)$ are easy to solve, since they are just a system of 2 (non-linear) equations for $p,n$ and $\Xi^{\pm}$, along with a system of 4 equations for the set $\Lambda,\Sigma^{\pm,0}$.
Once the Fermi momenta are known, Eqs.~(\ref{HF-eqs-1}) determine $e(k,f)$ at $T=0$. That is, at $T=0$ the HF equations for $e(k,f)$ do not require a self-consistent solution.

Having found an exact solution at $T = 0$, the iteration procedure described above works well.

%%%%%%%%%%%%%%%%%%%%%%%%%%%%%%%%%%%%%%%%%%%%%%
\subsection{Equations for the meson fields}
The full partition function is 
\[
Z=\int d\bar{\sigma}d\bar{\omega}d\bar{b}_{3}\,e^{-\beta(E_{meson}+\Phi_{B})} \, .
\]
To integrate over $\bar{\sigma},\bar{\omega},\bar{b}_{3}$ we use
the saddle point approximation. This amounts to estimating the integral
according to 
\[
Z\sim e^{-\beta(E_{meson}+\Phi_{B})_{saddle}} \, ,
\]
where an irrelevant multiplicative factor has been ignored and the saddle point is defined by 
\begin{equation}
\frac{d}{d\bar{\sigma}}\left(E_{meson}+\Phi_{B}\right)=0 \, ,
\label{eq:saddle-point}
\end{equation}
with analogous equations for the other mesons. They are determined
by the saddle point equations
\begin{eqnarray}
\frac{d}{d\bar{\sigma}}\left(E_{meson}+\Phi_{B}\right) & = & 0\nonumber \\
\frac{d}{d\bar{\omega}}\left(E_{meson}+\Phi_{B}\right) & = & 0
\label{MF_gen}\\
\frac{d}{d\bar{b}_{3}}\left(E_{meson}+\Phi_{B}\right) & = & 0 \, .
\nonumber 
\end{eqnarray}
Since $\Phi_{B}$ is stationary with respect to $e(k,f)$, because of Eqs.~(\ref{eq:HF_condition}), we do not need to worry about the dependence of $\Phi$ on $\bar{\sigma},\bar{\omega},\bar{b}_{3}$ through $e(k,f)$. 
On the other hand, we must take into account the dependence of $W$
on the $\sigma$ field. We note that this rearrangement effect, arising from the field dependence of the interaction, was omitted in Refs.~\cite{Stone2007,Stone2021}. 

From (\ref{eq:Phi-generic}) we find
\begin{eqnarray*}
\frac{d}{d\bar{\sigma}}\Phi_{B} & = & \frac{2{\cal V}}{(2\pi)^{3}}\sum_{f}\int d\vec{k}n(k,f)\frac{d}{d\bar{\sigma}}K(k,f)+\\
 &  & \frac{{\cal V}}{(2\pi)^{6}}\sum_{ff'}\int d\vec{k}d\vec{q}n(k,f)n(q,f')\frac{dW^{ff'}(\vec{k},\vec{q})}{d\bar{\sigma}} \, , 
\end{eqnarray*}
with analogous equations for $\bar{\omega},\bar{b}_{3}$. Then, using Eqs.~(\ref{eq:K},\ref{eq:Emeson}), we obtain the mean field equations:
\begin{eqnarray}
\frac{dV(\bar{\sigma})}{d\bar{\sigma}}+\frac{2}{(2\pi)^{3}}\sum_{f}\int d\vec{k} \, n(k,f)\frac{d}{d\bar{\sigma}}K(k,f)+\nonumber \\
\frac{1}{(2\pi)^{6}}\sum_{ff'}\int d\vec{k} \, d\vec{q} \, n(k,f) \, n(q,f') \frac{dW^{ff'}(\vec{k},\vec{q})}{d\bar{\sigma}} & = & 0\nonumber \\
-m_{\omega}^{2} \, \bar{\omega}+\frac{2}{(2\pi)^{3}}\sum_{f}g_{\omega}^{f}\int d\vec{k} \, n(k,f) & = & 0\label{Mean_fields}\\
-m_{\rho}^{2} \, \bar{b}_{3}+g_{\rho}\frac{2}{(2\pi)^{3}}\sum_{f}m(f)\int d\vec{k}\, n(k,f) & = & 0 \, . 
\nonumber 
\end{eqnarray}

In summary, the calculation of the grand potential $\Phi=\Phi_{B}+E_{meson}$
involves:
\begin{itemize}
\item the solution of the self-consistent HF equations, 
Eqs.~(\ref{HF-eqs}), to
determine the single particle energies, $e(k,f)$
\item the solution of the mean field equations, Eqs.~(\pageref{Mean_fields}), 
to determine $\bar{\sigma},\, \bar{\omega}, \, \bar{b}_{3}$ -- see the Appendix
for details.
\item the calculation of $\Phi_{B}$ according to 
Eq.~(\ref{Phi_B1}) or (\ref{Phi_B2})
\end{itemize}
After substitution of the solutions $e(k,f), \bar{\sigma}, \bar{\omega}$ 
and $\bar{b}_{3}$,
$\Phi$ becomes a function of $\mu$ and $\beta$ and we can compute the
useful thermodynamic quantities. 

%%%%%%%%%%%%%%%%%%%%%%%%%%%%%%%%%%%%
\section{Thermodynamic quantities}
\label{sec:thermo}
%\subsection{Particle number}
From the definition of the partition function we have 
\[
\langle N(f)\rangle=- \, \frac{\partial\Phi}{\partial\mu(f)} 
\]
and using Eqs.~(\ref{HF-eqs}) one readily finds the following relations 
%(see Appendix 4)
% 
\begin{eqnarray}
\frac{\partial\Phi}{\partial\mu(f)} & = & \frac{\partial\Phi_{B0}}{\partial\mu(f)}\label{eq:Usefull}\\
\frac{\partial\Phi}{\partial\beta} & = & \frac{\partial\Phi_{B0}}{\partial\beta} \, . 
\nonumber 
\end{eqnarray}
This implies the simple result 
\begin{equation}
\langle N(f)\rangle=-\frac{\partial\Phi_{B0}}{\partial\mu(f)}=\frac{2{\cal V}}{(2\pi)^{3}} %\sum_{f}
\int d\vec{k} \, n(k,f) \, .
\label{eq:part_number}
\end{equation}
Thus, because of the self-consistency conditions, we recover the naive
expression for the particle number, which must be computed with the self-consistent
energy $e(k,f)$, rather than the mean field energy $K(k,f)$.

%%%%%%%%%%%%%%%%%%%%%%%
\subsection{Entropy}
 Using relation (\ref{eq:Usefull}) we have
\[
\frac{\partial}{\partial\beta}\Phi=\frac{\partial}{\partial\beta}\Phi_{B0}=-\frac{1}{\beta}\Phi_{B0}+\frac{1}{\beta}\frac{2{\cal V}}{(2\pi)^{3}}\sum_{f}\int d\vec{k}\, n(k,f)\left(e(k,f)-\mu(f)\right) \, , 
\]
and hence
\begin{eqnarray*}
\langle S\rangle & = & \frac{\partial}{\partial T}T\ln Z=\beta^{2}\frac{\partial}{\partial\beta}\Phi=\beta^{2}\frac{\partial}{\partial\beta}\Phi_{B0}\\
 & = & {\cal V}\beta\frac{2}{(2\pi)^{3}}\sum_{f}\int d\vec{k}\left[\frac{1}{\beta}\ln\left(1+e^{-\beta\left(e(k,f)-\mu(f)\right)}\right)+n(k,f)\left(e(k,f)-
\mu(f)\right)\right] \, . 
\end{eqnarray*}
Using this expression one can check that $S\to0$ when $T\to0$,
as it should. 

%%%%%%%%%%%%%%%%%%%%%%%%%%
\subsection{Energy}
From the defining relation 
\[
\langle E-\mu N\rangle=-\frac{\partial}{\partial\beta}\ln Z=\Phi+\beta\frac{\partial}{\partial\beta}\Phi \, , 
\]
we have:
\begin{eqnarray*}
\langle E-\mu N\rangle & = & \Phi-\Phi_{B0}+\frac{2{\cal V}}{(2\pi)^{3}}\sum_{f}\int d\vec{k}n(k,f)\left(e(k,f)-\mu(f)\right)\\
 & = & E_{mesons}+\frac{{\cal V}}{(2\pi)^{3}}\sum_{f}\int d\vec{k}n(k,f)\left[K(k,f)-e(k,f)\right]+\\
 &  & \frac{2{\cal V}}{(2\pi)^{3}}\sum_{f}\int d\vec{k}n(k,f)\left(e(k,f)-
\mu(f)\right) \, , 
\end{eqnarray*}
so
\begin{equation}
\langle E\rangle=E_{mesons}+\frac{{\cal V}}{(2\pi)^{3}}\sum_{f}\int d\vec{k}n(k,f)\left[K(k,f)+e(k,f)\right] \, . 
\label{eq:energy}
\end{equation}
It is obvious that the thermodynamic relation 
\[
\langle E\rangle=\Phi+T\langle S\rangle+\mu\langle N\rangle=-P+T\langle S\rangle+\mu\langle N\rangle
\]
is satisfied by the above expressions and we use it to compute the
energy.

%%%%%%%%%%%%%%%%%%%%%%%%%%%%%%%%%%%%%%%%%%%%
\subsection{$\beta$ equilibrium equations}
\label{sec:betaeq}
    
The chemical potentials are in equilibrium under the constraint of
(local) electric and baryon charge conservation and the antiparticles
must satisfy the relation $\mu(\bar{a})=-\mu(a)$. Then, if the lepton numbers ($L_{e},L_{\mu}$) are conserved, all lepton species $(e^{-},\bar{\nu}_{e},e^{+},\nu_{e})$ and $(\mu^{-},\bar{\nu}_{\mu},\mu^{+},\nu_{\mu})$, 
must be present. This corresponds to the neutrino trapping case. On
the other hand, if the neutrinos escape from the (proto-)star,  then the lepton number is not conserved. Hence we distinguish 2 cases.

%%%%%%%%%%%%%%%%%%%%%%%%%%%%%%%%%%%%%%%%%%%%
\subsubsection{Lepton number is not conserved}
\label{sec:chempot}
Assuming that there are no trapped neutrinos, we have the equilibrium equations:
\begin{eqnarray*}
\mu(\mu^{-}) & = & \mu(e^{-})\\
\mu_{i}-\mu_{n} & = & -\mu(e^{-})Q_{i},\,\,i=p,\Lambda,...
\end{eqnarray*}
and the conservation equations: 
\begin{eqnarray*}
n(e^{+})-n(e^{-})+n(\mu^{+})-n(\mu^{-}) & = & -\sum_{i}\rho_{i}Q_{i}\\
\sum_{i}\rho_{i} & = & \rho_{B} \, , 
\end{eqnarray*}
where $\rho_{i}$ are the baryon densities. If we assume that the
positive leptons are present, we use $\mu(e^{+},\mu^{+})=-\mu(e^{-},\mu^{-})$
to compute their densities.

%%%%%%%%%%%%%%%%%%%%%%%%%%%%%%%%%%%%%%%
\subsubsection{Lepton number is conserved}
In this case all neutrinos and charged leptons are active, so the equilibrium equations
become
\begin{eqnarray}
%j7m,,ii,ui...
\mu(e^{-})-\mu(\nu_{e}) & = & \mu(\mu^{-})-\mu(\nu_{\nu})\\
\mu_{i}-\mu_{n} & = & -\left[\mu(e^{-})-\mu(\nu_{e})\right]Q_{i} \, , 
\end{eqnarray}
while the conservation equations are:
\begin{eqnarray}
n(e^{-})+n(\nu_{e})-n(e^{+})-n(\bar{\nu}_{e}) & = & L_{e}\\
e\to\mu & = & L_{\mu}\\
n(e^{+})-n(e^{-})+n(\mu^{+})-n(\mu^{-}) & = & -\sum_{i}\rho_{i}Q_{i}\\
\sum_{i}\rho_{i} & = & \rho_{B} \,. 
\end{eqnarray}
To determine the equilibrium composition completely one must specify the lepton numbers (or fractions).

%%%%%%%%%%%%%%%%%%%%%%%%%%%%%%%%%%%%%%%%%
\section{Computation method}
\label{sec:com}

To demonstrate the effect of the Fock terms in the QMC energy
functional, we need to check that their effect can be distinguished from the consequence of choosing different input parameters in a simple mean-field treatment.
Consistent with our previous work~\cite{Stone2021,Stone2022c}, we used the QMC model with five
parameters, three meson-nucleon coupling constants, G$_\sigma$,
G$_\omega$ and G$_\rho$, as well as the mass of the $\sigma$ meson, $M_\sigma$, and
the strength of the sigma meson cubic self-coupling, $\lambda_3$. 
As before,
$M_\sigma$ was fixed to 700 MeV and
$\lambda_3$ to zero, leaving  the three coupling constants
variable. These couplings are adjusted in the model to reproduce the
empirical values of parameters of symmetric nuclear matter (SNM) at saturation
 density, $\rho_0$, the energy per particle, E$_0$/A, and  the symmetry
energy coefficient, $J$, of asymmetric nuclear matter
(ANM). To our knowledge,
such a mapping of the QMC and nuclear matter parameter spaces, being of
the same dimension, is a unique feature of the QMC model.
However, $\rho_0$,  E$_0$/A and  $J$ are correlated and not exactly
known (see e.g.,~\cite{Stone2021a,Stone2022c} and their
range is a subject of ongoing research 
(see for example Ref.~\cite{Horowitz2014}).

In order to identify ranges of the QMC coupling constants, compatible
with the generally acceptable ranges of the SNM parameters, we have adopted
the method used in Ref.~\cite{Stone2022c}. Since then the QMC-A
model has been further technically developed and the results changed in
a minor way, as discussed later. Also, the previous work focused on
the role of the symmetry energy in the high-density matter in
astrophysical objects, whereas in the present work we investigate the role
of the exchange terms in the QMC model.

Here we construct a 3D rectangular mesh with sides
$\rho_0$=0.14-0.18 fm$^{-3}$ (in steps of 0.01 fm$^{-3}$),  $E_0$/A=-14 to -18 MeV (in steps of 1 MeV) and $J$ between 28 and 32 MeV (in steps of 2 MeV, a total of 110 points. At each
point, the QMC predictions for the slope of the
symmetry energy $L$, the volume incompressibility $K$, the couplings
constants G$_\sigma$,  G$_\omega$,  G$_\rho$, and the single-particle
potentials U$_Y$ for Y=$\Lambda$, $\Sigma$ and $\Xi$
hyperons at saturation density in symmetric nuclear matter are computed.  For each of these choices we compute the gravitational mass, radius
and central density of a maximum mass hot proto-neutron star (PNS) with
a fixed entropy density, S/A=2k$_B$, as well as the radius, central density
and tidal deformability of a cold 1.4 M$_\odot$ neutron star. These
calculations were performed both with (HF - Hartree-Fock) and without (MF - mean-field) the
exchange Fock term in the QMC calculation, under exactly the same
thermodynamic conditions. In this way it was possible to compare
results with and without the inclusion of the exchange terms and
thus eliminate any ambiguity in identifying the effect of the Fock terms. 
\begin{figure}
  \includegraphics[width = 0.75\textwidth]{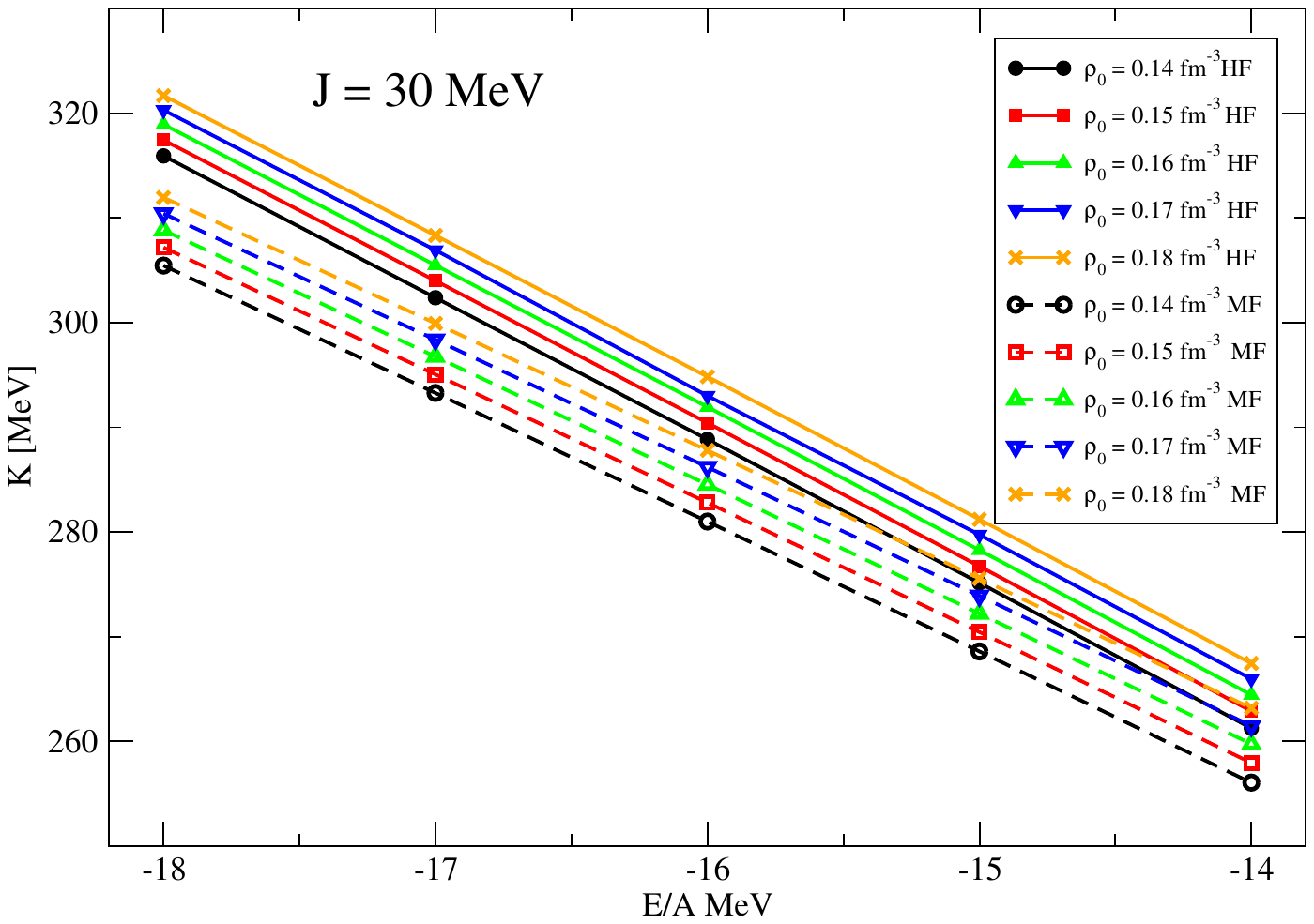}
  \caption{ Volume incompressibility as a function of E$_0$/A for
    fixed  $\rho_0$ in the range 0.14 to 0.18 fm$^{-3}$ and J=30
    MeV. Solid (dashed) curves and full (empty) symbols represent
    calculation with (without) HF (MF) the exchange term. }
 \label{fig:1}
\end{figure}
%

%%%%%%%%%%%%%%%%%%%%%%%%%%%%%%%%%%%%%%%%%
\section{Results and Discussion}
\label{sec:resdis}

There are several possible paths one could follow in order to explore the sensitivity of the 
calculated observables to the selection of the planes cut through the
3D mesh of input parameters. In most cases, we have found that the most illustrative approach was to follow the 
dependence of an observable on E$_0$/A for fixed values of $\rho_0$ and $J$. While the sensitivity to $\rho_0$ was usually very telling, the sensitivity to the choice of $J$ was very limited and the resulting changes did not exceed a few
percent. Thus, for clarity of the figures, in most cases, we chose
J=30 MeV for illustration of the results. We note that a similar effect was already observed in Ref.~\cite{Stone2022c}.
\begin{figure}
  \includegraphics[width = 0.75\textwidth]{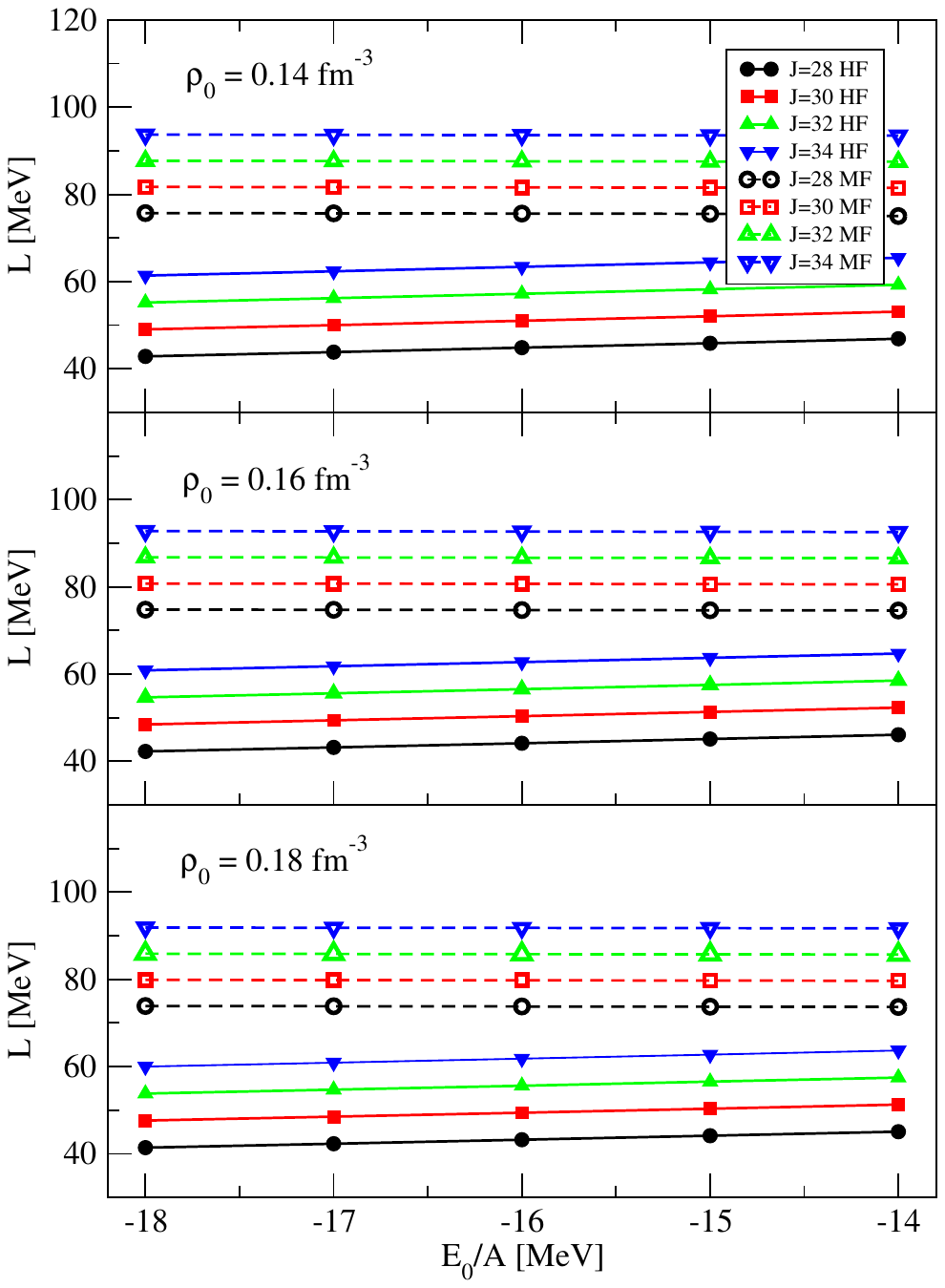}
  \caption{ Slope of the symmetry energy $L$ as a function of E$_0$/A for
    fixed  $\rho_0$ for 0.14, 0.16 and 0.18 fm$^{-3}$ and the range of
    J=28 to 34 MeV. Solid (dashed) curves and full (empty) symbols represent
    calculation with (without) HF (MF) the exchange term. }
 \label{fig:2}
\end{figure}
%

%A spreadsheet with all the results is available as Supplemental
%Material to this paper and the interested %reader can use that information to explore %other cuts through the parameter space. 
%
\begin{figure}
  \includegraphics[width = 0.75\textwidth]{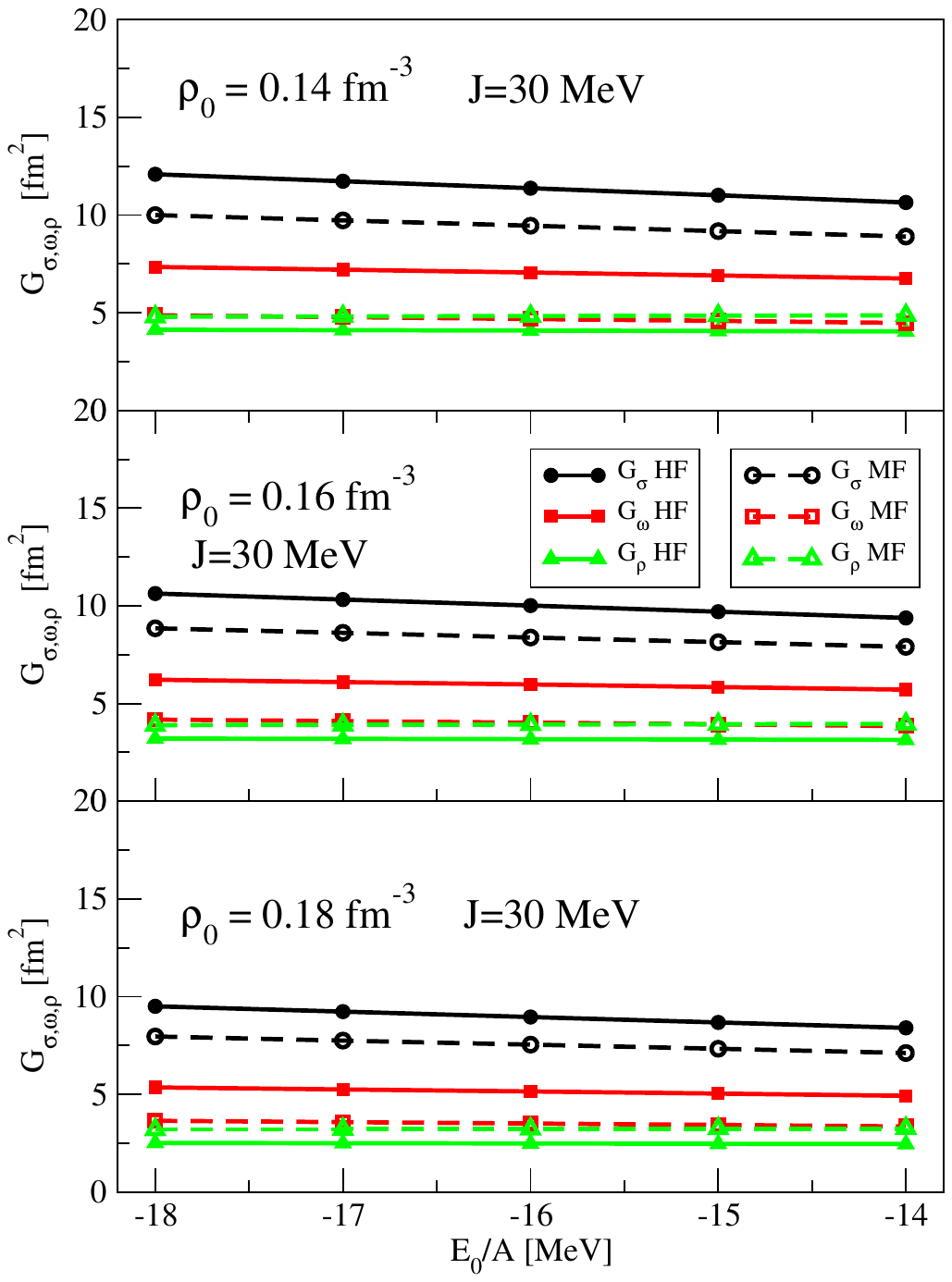}
  \caption{ QMC coupling constants G$_\sigma$,  G$_\omega$,  G$_\rho$  as a function of E$_0$/A for
    fixed  $\rho_0$ for 0.14, 0.16 and 0.18 fm$^{-3}$ and fixed J=30 MeV. Solid (dashed) curves and full (empty) symbols represent results calculated with (without) Fock terms. }
 \label{fig:3}
\end{figure}

Starting with Fig.~\ref{fig:1}, we observe the incompressibility K
decreasing linearly from 321.7 MeV at (E$_0$/A;$\rho_0$)=(-18 MeV;0.18 fm$^{-3}$) to
261.3 MeV at (E$_0$/A;$\rho_0$)=(-14 MeV;0.14 fm$^{-3}$) in the HF
calculation. The same pattern, with a minor difference in slope, is observed
in the MF model, with range of K values, from 312.0 MeV at
(E$_0$/A;$\rho_0$)=(-18 MeV;0.18 fm$^{-3}$) to 256.0 MeV at
(E$_0$/A;$\rho_0$)=(-14 MeV;0.14fm$^{-3}$). The incompressibility is systematically lower in the MF model, except for a minor overlap where $\rho_0$ takes its lowest values in the HF case and its highest values in the MF case.

The scenario illustrating the sensitivity of a quantity to
the variation of the symmetry energy parameter $J$ is shown in Fig.~\ref{fig:2}.  
We observe a clear separation of the HF results, exhibiting a weak
increase with E$_0$/A and $\rho_0$ and being lower in all cases than the MF results; the latter showing an even weaker opposite trend and being always
above the HF numbers. This behaviour is an obvious consequence of
the exchange terms. 
\begin{figure}
  \includegraphics[width = 0.75\textwidth]{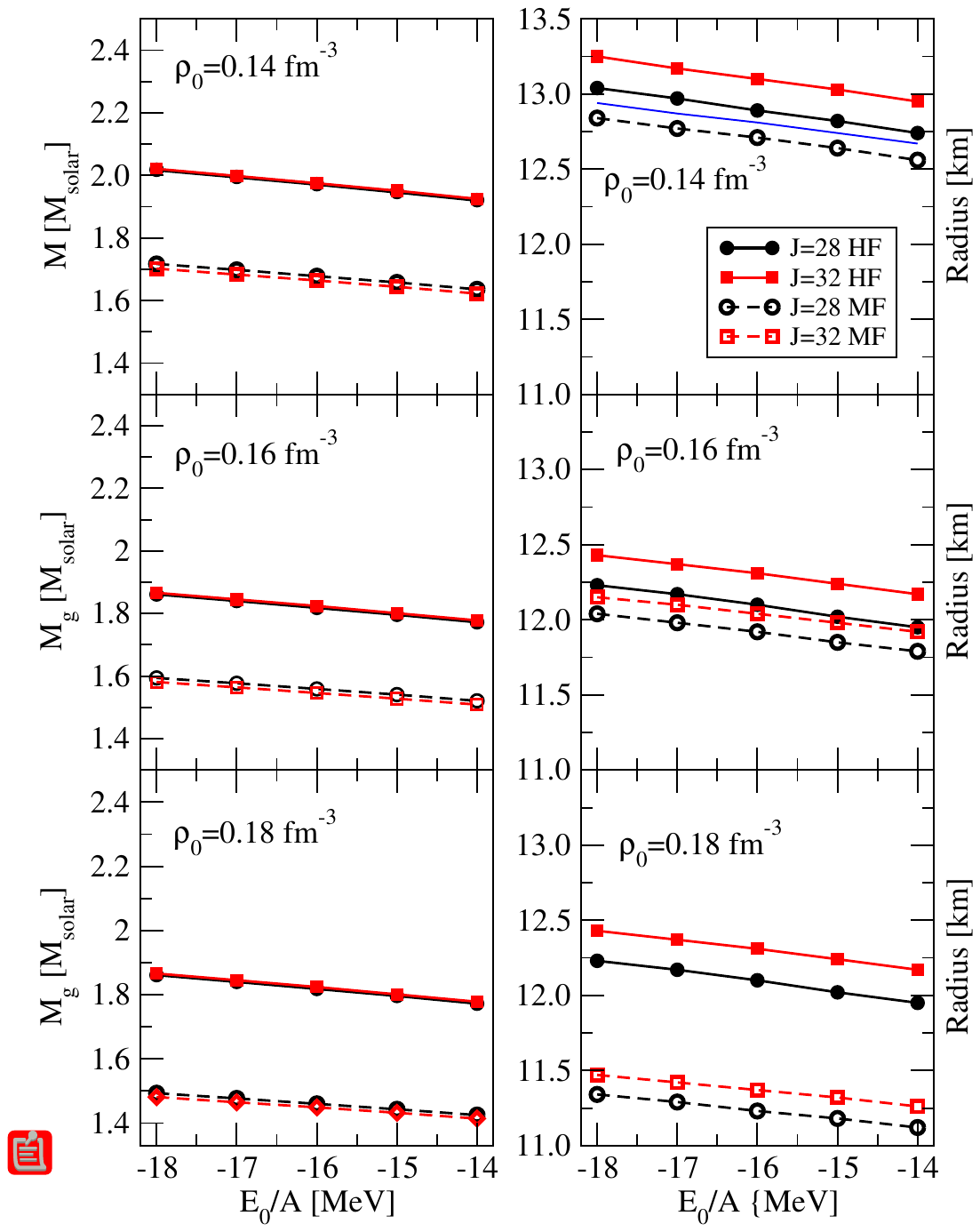}
  \caption{Maximum gravitational mass (left) and radius of that maximum mass star (right) at fixed entropy per baryon, S/A=2k$_B$, as a function of E$_0$/A for
    fixed symmetric nuclear matter parameters, $\rho_0$ equal to 0.14, 0.16 and 0.18 fm$^{-3}$ and fixed J=28
    and 32 MeV. Solid (dashed) curves and full (empty) symbols represent
    results calculated with (without) HF (MF) the exchange term. }
 \label{fig:4}
\end{figure}

The QMC coupling constants are obviously the main vehicle transporting
the exchange effects to the calculation of physical quantities at zero temperature. In
Fig.~\ref{fig:3} we show their sensitivity to these effects, in selected
scenarios which are further used to demonstrate the difference between the
HF and MF models. The couplings G$_\sigma$ and G$_\omega$ appear to be always
higher in the HF model, in contrast to G$_\rho$ which shows a minor
increase in the MF model above the HF model. Interestingly,
G$_\omega$ and G$_\rho$ are practically identical in the MF model
which is not the case in the HF model. Clearly, the coupling constants
are sensitive to the exchange terms, with G$_\sigma$ decreasing with
increasing $\rho_0$ and decreasing as E$_0$/A decreases in magnitude. These dependencies of the other constants are minor.

Turning now to compact objects, we explore the gravitational mass and
radius of a neutrinoless PNS with a core containing the full hyperon octet at constant entropy, S/A=2k$_B$.  Looking at the left panels of Fig.~\ref{fig:4}, two effects can be
observed. First, there is a difference between the maximum gravitational mass, M$_g$, 
that can be achieved in the HF and MF models, as much as 0.2
M$_\odot$. This is highly significant in this context. The effect of modest changes in $J$ in
both models is minimal and does not need to be taken into account. The
second effect is that of $\rho_0$, which reduces the maximum value of 
M$_g$ from about 2 M$_\odot$ at $\rho_0$=0.14 fm$^{-3}$ to about 1.8 M$_\odot$ at 0.18 fm$^{-3}$
but leaves the
difference between the HF and MF results very similar.

In the right hand panels, the effect of varying $\rho_0$ follows the trend
of the maximal mass, showing a reduction of the radius for decreasing
maximum mass. The difference between the HF and MF models is still
visible, showing systematically larger radii in the MF models. As before, the effect of the value of $J$ is minimal.
\begin{figure}
  \includegraphics[width = 0.75\textwidth]{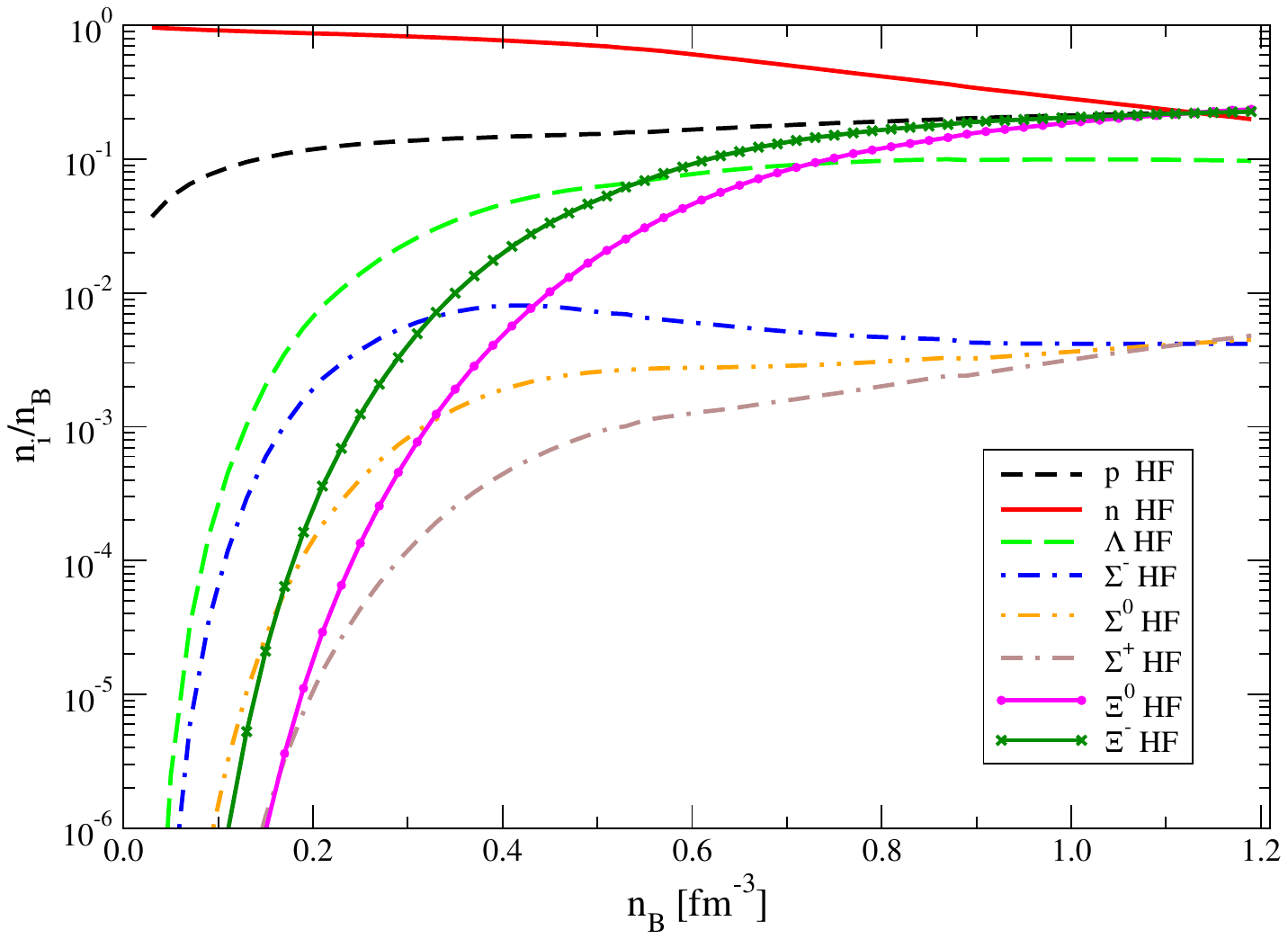}
  \caption{Relative population of nucleons and hyperons as a function
    of the total baryon density, $n_B$, as calculated in the HF model at fixed entropy per baryon, S/A=2k$_B$. Only
    population fractions higher than 10$^{-6}$ are shown.}
 \label{fig:5}
\end{figure}

\begin{figure}
  \includegraphics[width = 0.80\textwidth]{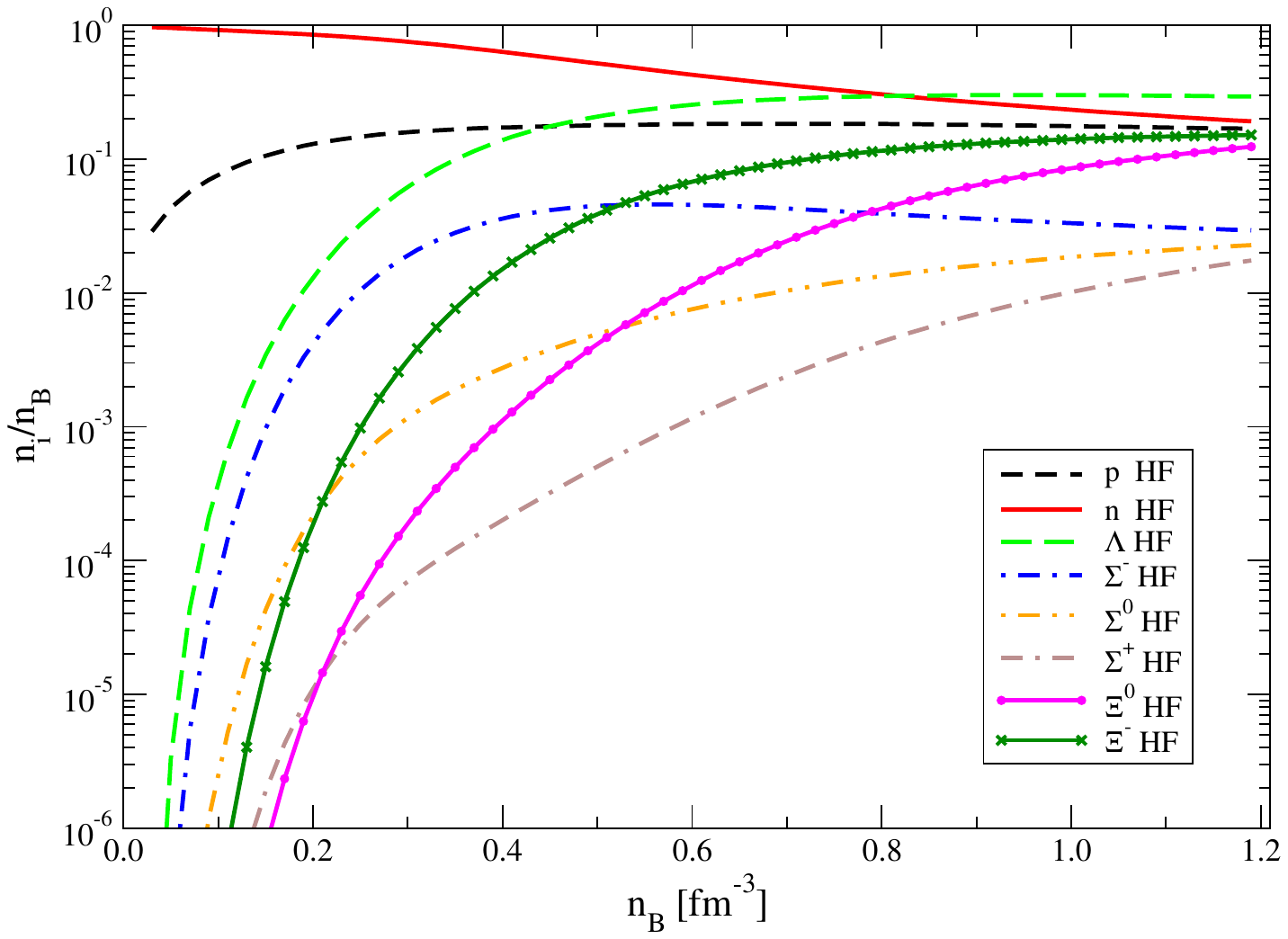}
  \caption{The same as Fig.~\protect\ref{fig:5} but for the MF model.}
 \label{fig:6}
\end{figure}

Taking the PNS stars with maximum mass, as indicated in top left panel in
Fig.~\ref{fig:4}, we show in Figs.~\ref{fig:5} and \ref{fig:6} (once again at fixed entropy per baryon, S/A=2k$_B$) the composition of the
PNS core as predicted by the HF and MF models. The effect of the exchange term is rather remarkable,
especially on the population of $\Sigma$ hyperons, showing a dramatic
decrease in population with increasing particle number density when Fock terms are included. Because at 
finite T hyperons are present at some level at all densities, there no density threshold for
their appearance. However, we can, for example, compare the fractional occurrence of some of the hyperons at a given baryon density. For example, the population of $\Xi^0$
baryons reaches $n_i/n_B$ = 10$^{-2}$, at a considerably higher density in the MF case.    

The appearance of hyperons in high density matter is
determined by the single-particle potentials, U$_Y$, which are dependent on
the nucleon-hyperon and hyperon-hyperon interactions. In cold matter,
they affect the density dependence of the threshold for hyperon
appearance, and in matter at finite temperature the hyperonic
population in the whole density spectrum.
\begin{figure}
  \includegraphics[width = 0.75\textwidth]{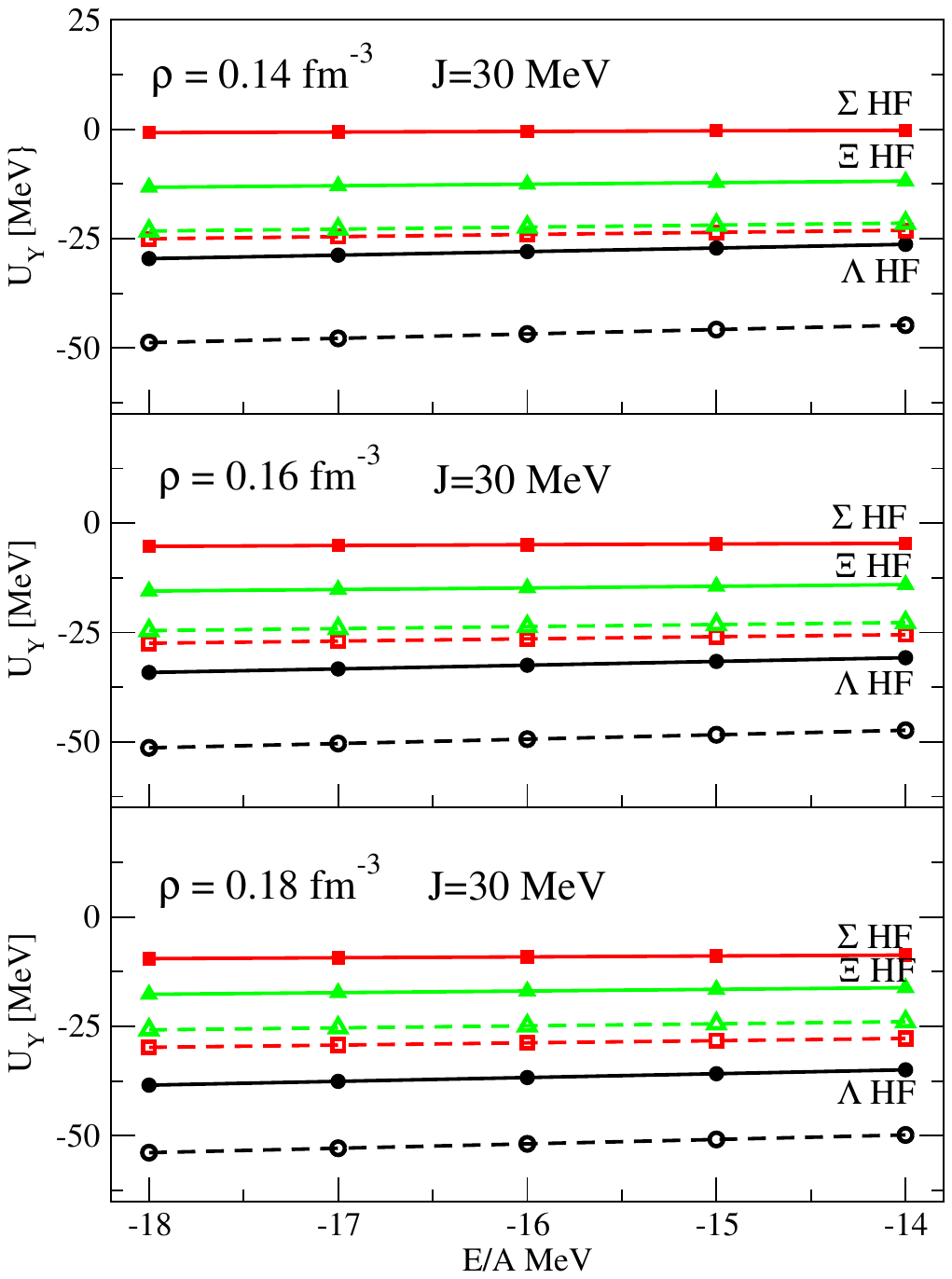}
  \caption{Hyperonic single-particle potentials (at T=0) versus E$_0$/A for $\rho_0$ equal to 0.14, 0.16 and 0.18 fm$^{-3}$ and  J=30 MeV.
   Only the HF results (solid curves and full symbols) are labeled in the graph, while the MF data are
    distinguished by dashed curves and empty symbols.}
 \label{fig:7}
\end{figure}

The potentials are treated as variable parameters in traditional RMF models but
appear naturally in the QMC model and thus are sensitive to the Fock terms. In particular, the appearance of $\Sigma$ hyperons
in the cores of cold neutron stars has been an issue for many years
(see e.g., Refs.~\cite {Providencia2019,Stone2022c} for more detail). We
illustrate the exchange term dependence of the U$_Y$ in 
Fig.~\ref{fig:7} for the $\Lambda, \Sigma$ and $\Xi$ hyperons. The
largest effect is observed for the $\Lambda$ hyperon and the smallest
for the $\Xi$ hyperon, but the difference between the HF and MF models
is clear. The dependence on $\rho_0$ and E$_0$/A is minor.
\begin{figure}
  \includegraphics[width = 0.75\textwidth]{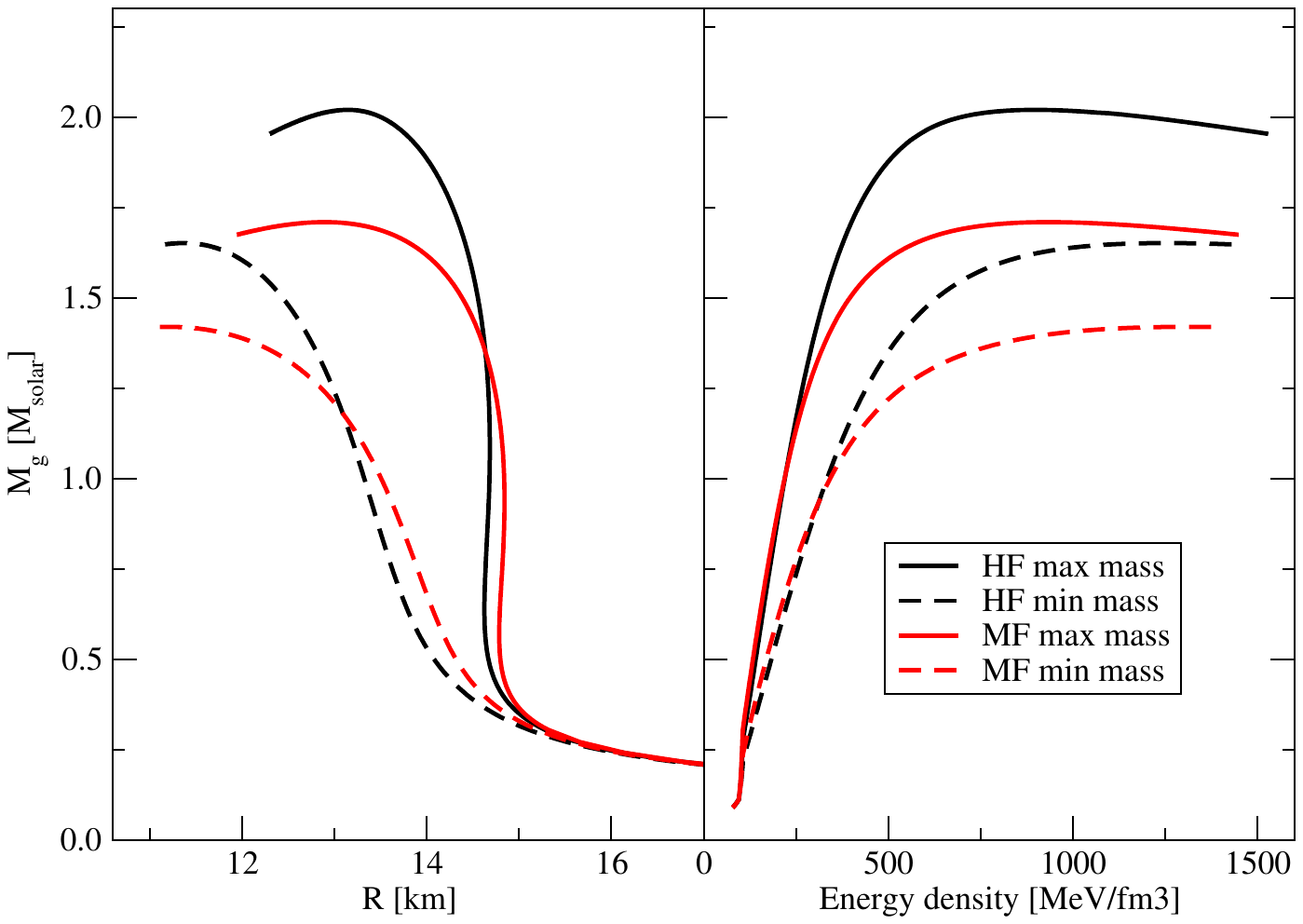}
  \caption{Gravitational mass vs radius of a PNS at fixed entropy per baryon, S/A=2k$_B$, for nuclear matter parameters that yield the global maximum
    and minimum mass. Left panel: Solid (dashed-dot) black curves represent M(R) curves for PNS with maximum (minimum) mass in the HF model. The red curves depict the same in the MF model. Right
    panel: The same as the left panel but for the mass dependence on
    the central energy density. For more explanation see text. }
 \label{fig:8}
\end{figure}

It is interesting to follow the global effect of exchange terms on the
mass-radius relation in warm stars with the full baryon octet in the
core. We selected
 PNSs with the maximum and minimum gravitational mass from all 109
 points within the input parameter space in both HF (black solid and
 dashed curves) and MF (red curves) models. The
 effect of the exchange terms is well demonstrated in
 gravitational masses, slopes of the curves and radii, as illustrated in Fig.~\ref{fig:8}, once again at fixed entropy per baryon, S/A=2k$_B$. It is notable that the central density reached in a maximum mass star is considerably higher in the MF case.
\begin{figure}
  \includegraphics[width = 0.75\textwidth]{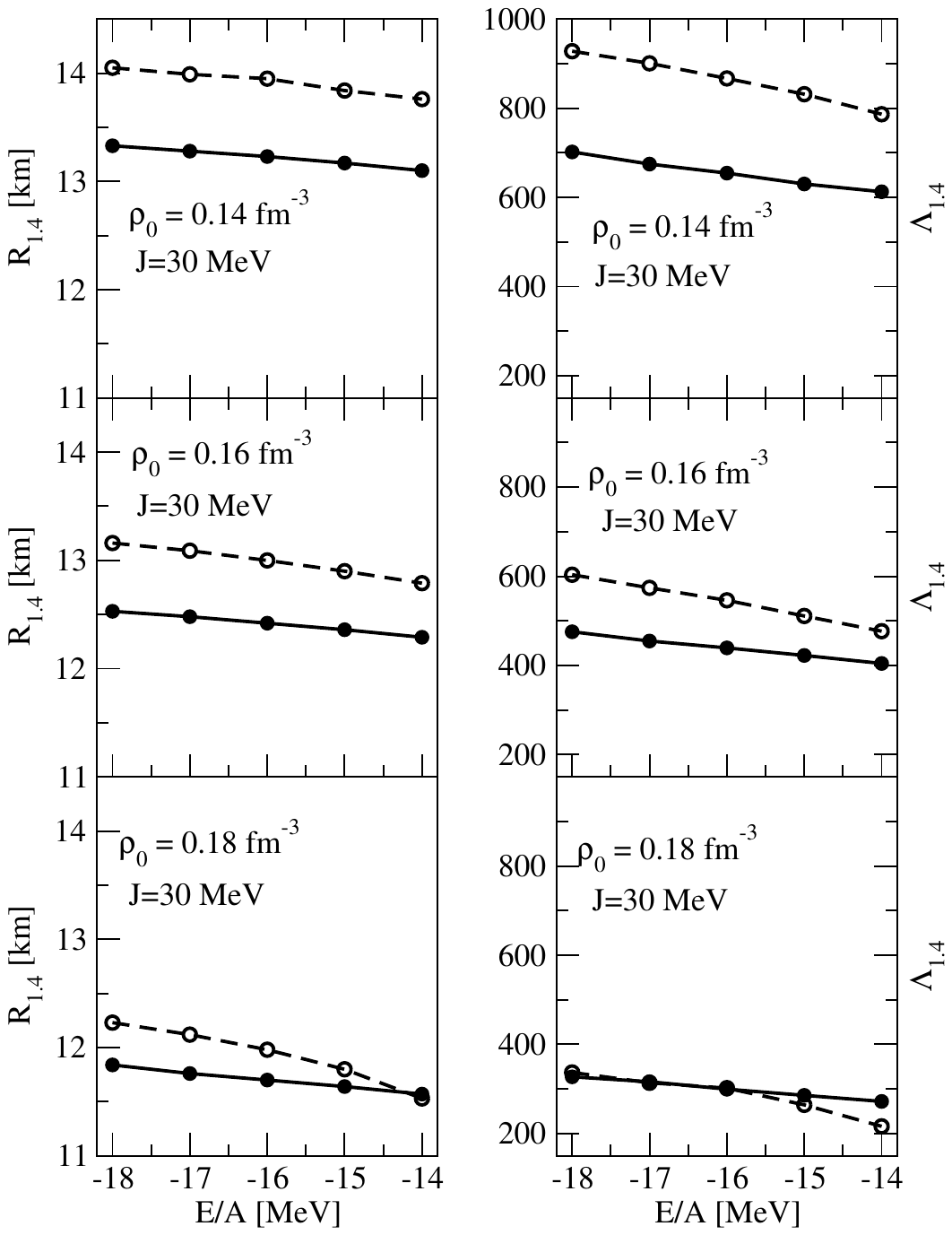}
  \caption{ R$_ {1.4}$ (left) and tidal deformability $\Lambda_{1.4}$
    (right)  of a cold neutron star vs E$_0$/A for
    fixed  $\rho_0$ for 0.14, 0.16 and 0.18 fm$^{-3}$ and J=30
    MeV cases. Solid (dashed) curves and full (empty) symbols represent
    results calculated with (without) HF (MF) the exchange term.  }
 \label{fig:9}
\end{figure}

Finally, we explore the prediction of the HF and MF models for the
radius, R$_ {1.4}$, and tidal deformability, $\Lambda_{1.4}$, of a cold 1.4 M$_{solar}$ star. This case is important because of the mergers of
neutron stars as a source of gravitational waves. Examination of
Fig.~\ref{fig:9} reveals the particular sensitivity to the choice of $\rho_0$
of the tidal deformability (see right panels), 
as well as the  
systematic decrease of the radius of the star as $\rho_0$ increases. At
the same time, the difference between the HF and MF models is quite
obvious in both quantities.    

%%%%%%%%%%%%%%%%%%%%%%%%%%%%%%%%%%%%%%%%%
\section{Concluding remarks}
\label{sec:concl}

We have presented the first complete formal development of a relativistic Hartree-Fock treatment of the EoS of dense matter at finite temperature including hyperons. The Fock terms are, of course, essential to ensure that the participating baryons obey the Pauli Exclusion Principle. The formalism was then used to explore the importance of including the Fock terms, as opposed to the much simpler application of mean-field theory.

In order to demonstrate the relevance of the Fock terms, it was essential to distinguish their effect from the choice of input parameters. This investigation confirmed and extended the findings reported in Ref.~\cite{Stone2022c}, namely that the input parameters do have important effects on the results. However, those parameters are not as well known as one would like. 

The first finding was that in almost all cases the properties of neutron stars were linearly dependent on the input parameters within their range of uncertainty. The most important property investigated was the maximum mass of the stars, for which the highest sensitivity corresponded to variations in $\rho_0$, while there was very little sensitivity to $J$. 

By exploring the properties of neutron stars across an extensive mesh of EoS calculated over a range of nuclear matter parameters, it was possible to establish clear differences in the predictions with and without the Fock terms; differences which cannot be mimicked by the variation of nuclear matter parameters within the generally accepted range of uncertainty. Figures~\ref{fig:5} and \ref{fig:6} illustrate the difference between the fractions of various hyperons in these two cases.

By far the most important difference is illustrated in Fig.~\ref{fig:8}, where we see that for those sets of nuclear matter parameters which produce the largest or smallest maximum mass stars in either the mean-field or Hartree-Fock cases, both the maximum and minimum values are considerably larger in the Hartree-Fock calculations and the radii of the stars are significantly smaller.

In the light of these results we suggest that it will be important for future theoretical studies of neutron star properties, for cold and especially warm stars, to include the Fock terms. 

%%%%%%%%%%%%%%%%%%%%
\section*{Acknowledgements}

JRS and PAMG are grateful for the hospitality at the University of Adelaide during some parts of this work. This work was supported in part by the University of Adelaide and by the Australian Research Council under the Discovery Project DP230101791 and through the ARC Centre of Excellence for Dark Matter Particle Physics.\\
https://www.overleaf.com/project/64f3b2010d0139d74fdb438d

\bibliography{Fock}

\end{document}